\documentclass[12pt]{article}
\usepackage{amssymb}
\usepackage{graphicx,psfrag}
\usepackage[nosort]{cite}

\newcommand{\bea}{\begin{eqnarray}}
\newcommand{\eea}{\end{eqnarray}}
\newcommand{\simgt}{\hbox{ \raise3pt\hbox to 0pt{$>$}\raise-3pt\hbox{$\sim$} }}
\newcommand{\simlt}{\hbox{ \raise3pt\hbox to 0pt{$<$}\raise-3pt\hbox{$\sim$} }}

\newcommand{\clfn}{\setcounter{footnote}{0}}

\begin{document}

\begin{titlepage}
\title{\bf \Large
\vspace{28mm}
Renormalon Cancellation and Perturbative QCD Potential
as a Coulomb+Linear Potential\thanks{
Invited talk given at the ``International Conference on
Color Confinement and Hadrons in Quantum Chromodynamics
(Confinement 2003)'', Riken, Tokyo, 21--24 July 2003.
}\vspace{7mm}}
\author{
Y.~Sumino
\\ \\ \\ Department of Physics, Tohoku University\\
Sendai, 980-8578 Japan
}
\date{}
\maketitle
\thispagestyle{empty}
\vspace{-4.5truein}
\begin{flushright}
{\bf TU--698}\\
{\bf October 2003}
\end{flushright}
\vspace{4.5truein}
\begin{abstract}
\noindent
{\small
Recently evidence has been found that the perturbative QCD potential
agrees well with phenomenological potentials and lattice computations of
the QCD potential.
We review the present status of the perturbative QCD potential and
theoretical backgrounds.
We also report our recent analysis which shows analytically, on the 
basis of renormalon dominance picture, that the
perturbative QCD potential quickly ``converges'' to a Coulomb-plus-linear
form.
The Coulomb-plus-linear potential can be computed systematically 
as we include more terms of the perturbative series;
up to three-loop running (our current best knowledge), 
it shows a convergence towards lattice results.
e.g.\ At one-loop running, the linear potential is $\sigma r$ with
$\sigma = (2\pi C_F/\beta_0)\Lambda_{\overline{\rm MS}}^2$.
}
\end{abstract}
\vfil

\end{titlepage}

\section{Introduction}
In this article, we review the perturbative QCD predictions of
the QCD potential
for a static quark-antiquark ($Q\bar{Q}$) pair,
in the distance region relevant to the bottomonium and charmonium
states, namely, 
$0.5~{\rm GeV}^{-1} (0.1~{\rm fm}) \simlt r \simlt
5~{\rm GeV}^{-1} (1~{\rm fm})$.

Several years ago, the perturbative prediction of the QCD potential
became much more accurate in this region.
There were two important developments:
(1) The complete ${O}(\alpha_S^3)$ corrections to the
QCD potential have been computed \cite{ps};
also the relation between the quark pole mass and the $\overline{\rm MS}$
mass has been computed up to ${O}(\alpha_S^3)$ \cite{mr}.
(2) A renormalon cancellation was discovered
\cite{renormalon} in 
the total energy of a static 
$Q\bar{Q}$ pair\footnote{
By ``static'', we mean that the kinetic energies of $Q$ and $\bar{Q}$
are neglected.
},
$E_{\rm tot}(r) \equiv 2 m_{\rm pole} + V_{\rm QCD}(r)$.
Consequently, convergence of the
perturbative series of $E_{\rm tot}(r)$
improves drastically,
if it is expressed in terms of the quark $\overline{\rm MS}$ mass
instead of the pole mass.

\begin{figure}[t]\centering
  \psfrag{muGeV}{\small\hspace{-3mm}$\mu$~[GeV]}
  \psfrag{VGeV}{\small\hspace{-7mm}$E_{\rm tot}-2m$~~[GeV]}
  \psfrag{r=2r0}{\small\hspace{-20mm}$r=2\,r_0\approx 1$~fm}
  \psfrag{polemass}{\small\hspace{-10mm}\raise-0pt\hbox{Pole-mass scheme}}
  \psfrag{MSbar}{\small\hspace{-8mm}$\overline{\rm MS}$-mass scheme}
  \psfrag{m=3}{\small\hspace{-5mm}($\overline{m}=3$~GeV)}
  \includegraphics[width=15cm]{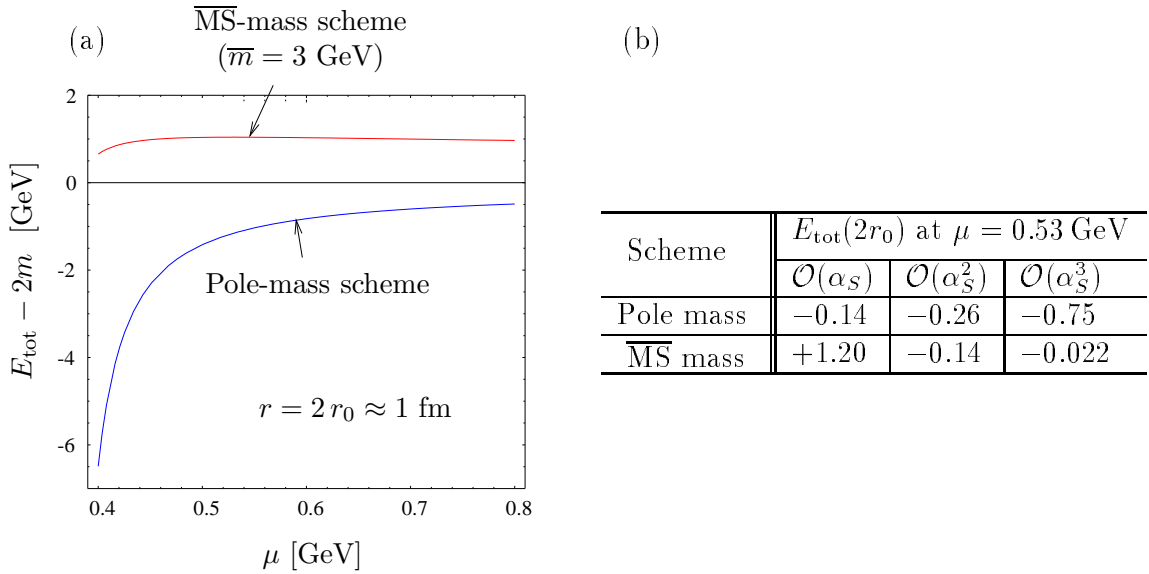}
\caption{\small (a)Scale dependences of $E_{\rm tot}(r)$ at 
$r =2r_0 \approx 1~{\rm fm}$, in the pole-mass and 
$\overline{\rm MS}$-mass schemes.
($\overline{m}\equiv m_{\overline{\rm MS}}( m_{\overline{\rm MS}})
= 3$~GeV)
(b) Perturbative series of $E_{\rm tot}(r)$ at $r =2r_0$,
and at $\mu = 0.53$~GeV determined from 
$\partial E_{\rm tot}/\partial \mu = 0$ in (a) 
($\overline{\rm MS}$-mass scheme).
\label{Etot-comb}}
\end{figure}

Let us demonstrate the improvement of accuracy of the perturbative
prediction for $E_{\rm tot}(r)$ up to ${O}(\alpha_S^3)$,
in the case without light quark flavors ($n_l=0$).
In Fig.~\ref{Etot-comb}(a), we fix $r$ to be $2r_0 \approx 1~{\rm fm}$
as an extreme long-distance case\footnote{
$r_0\approx 0.602\, \Lambda_{\overline{\rm MS}}^{-1} $ 
\cite{Capitani:1998mq} represents the Sommer scale.
It is translated to physical unit
following the convention of lattice
calculations in the quenched approximation:
$r_0 \approx 0.5$~fm.
} 
and show
the renormalization-scale ($\mu$) dependence of
$E_{\rm tot}(r=2r_0)$.
We see that $E_{\rm tot}$ is much less scale dependent when we
use the $\overline{\rm MS}$ mass instead of the pole mass.
Fig.~\ref{Etot-comb}(b) compares the convergence behaviors
of the perturbative series of $E_{\rm tot}$ for the same $r$
and when $\mu$ is fixed to the value where $E_{\rm tot}$
becomes least sensitive to variation of $\mu$ 
(minimal-sensitivity scale).
If we use the pole mass, the series is divergent, whereas
for the $\overline{\rm MS}$ mass, the series is convergent.\footnote{
When the series is convergent
($\overline{\rm MS}$-mass scheme), $\mu$-dependence decreases
as we include more terms of the perturbative series
into $E_{\rm tot}(r)$,
whereas when the series is divergent
(pole-mass scheme), $\mu$-dependence does not decrease with
increasing order.
}
In fact, we observe qualitatively the same features at
different $r$ and for different number of light quark
flavors $n_l$.
Generally, at smaller $r$,
$E_{\rm tot}(r)$ becomes less
$\mu$-dependent and more convergent.

The aim of this paper is to study properties of $E_{\rm tot}(r)$,
given the much more accurate prediction as compared to several
years ago.
In Sec.~2 we examine $E_{\rm tot}(r)$ up to $O(\alpha_S^3)$.
Sec.~3 provides our present theoretical understanding based on
renormalon dominance picture and operator-product-expansion (OPE).
We show that the perturbative QCD potential ``converges'' 
to a Coulomb+linear potential in Sec.~4.
Conclusions are given in Sec.~5.

\section{\boldmath $E_{\rm tot}(r)$ up to $O(\alpha_S^3)$}
\clfn

The most solid part of our study is an examination of the
perturbative predictions of $E_{\rm tot}(r)$ up to $O(\alpha_S^3)$,
through comparisons 
with phenomenological potentials and with lattice computations
of the QCD potential.

\begin{figure}[t]\centering
  \hspace*{-22mm}
  \includegraphics[width=19.2cm]{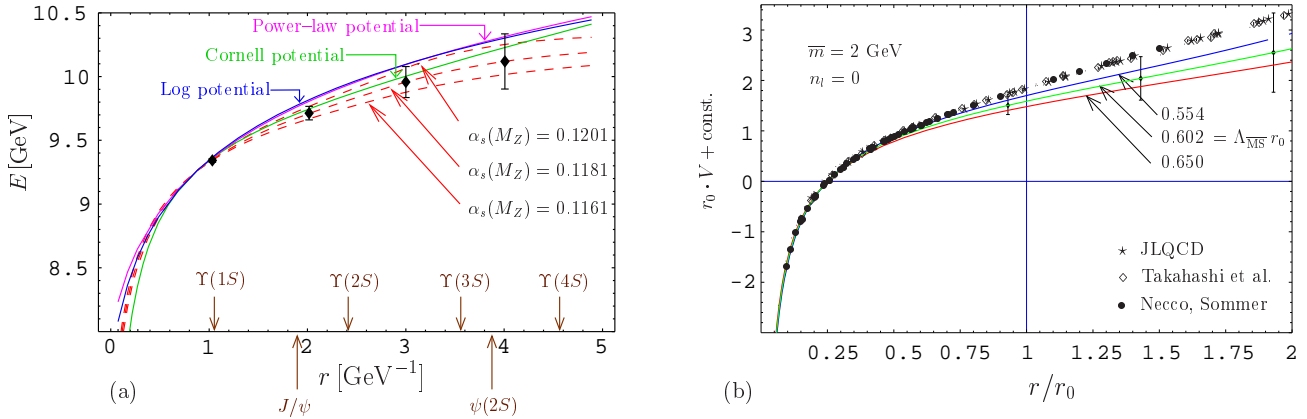}
\caption{\small (a) Comparison between $E_{\rm tot}(r)$ and typical
phenomenological potentials \cite{Recksiegel:2001xq}.
Arrows at the bottom show the r.m.s.\ radii of the heavy
quarkonium states.
(b) Comparison between $E_{\rm tot}(r)$ and lattice
computations of the QCD potential \cite{RS-lat}.
\label{comppot}}
\end{figure}

In Fig.~\ref{comppot}(a), $E_{\rm tot}(r)$ is compared with typical
phenomenological potentials.
Since the latter are determined from
the heavy quarkonium spectra, we choose realistic values for
the input parameters of $E_{\rm tot}(r)$:
$\overline{m}_b = 4.190$~GeV;
$n_l=4$ with $\overline{m}_u=\overline{m}_d=\overline{m}_s=0$
and $\overline{m}_c=1.243$~GeV for the light quarks in internal loops;
$\mu =\mu(r)$ is fixed by either
$\partial E_{\rm tot}/\partial \mu=0$ or
$|E_{\rm tot}^{(3)}|=\mbox{minimum}$, but both
prescriptions lead to almost same values of $E_{\rm tot}(r)$.
See \cite{Sumino:2001eh,Recksiegel:2001xq} for details.
We see that $E_{\rm tot}(r)$ corresponding to the present
values of the strong coupling constant (dashed lines)
agree well with the phenomenological potentials
within estimated perturbative uncertainties
(indicated by error bars).
We also note that the agreement is lost quickly if we take
$\alpha_S(M_Z)$ outside of the present world-average
values, so that the agreement is unlikely to be accidental.

In Fig.~\ref{comppot}(b), $E_{\rm tot}(r)$ is compared with
the recent lattice computations of the QCD potential
in the quenched approximation.
Accordingly we set $n_l=0$ in $E_{\rm tot}(r)$.
We take $\overline{m}=2$~GeV, which stabilizes the
perturbative prediction up to largest $r$.
The scale $\mu = \mu(r)$ is fixed in the same way as in Fig.~\ref{comppot}(a).
See \cite{RS-lat} for details.
The three solid lines in Fig.~\ref{comppot}(b) 
represent the same perturbative
prediction for $E_{\rm tot}(r)$;
there is an error in the relation between $\Lambda_{\overline{\rm MS}}$ and 
the lattice scale, and
the three lines represent the error band of this
relation $\Lambda_{\overline{\rm MS}} r_0 = 0.602(48)$
as given by \cite{Capitani:1998mq}.
Taking this uncertainty into account and also uncertainties of the
perturbative prediction (error bars), we observe a good agreement
between $E_{\rm tot}(r)$ and lattice results.

In fact, by now several works have confirmed the agreements 
\cite{Sumino:2001eh,necco-sommer,Recksiegel:2001xq,Pineda:2002se,RS-lat}.
Although details depend on how the renormalon
in the QCD potential is cancelled, qualitatively the same conclusions
were drawn, i.e.\ perturbative predictions become accurate
and agree with phenomenological potentials/lattice results
up to much larger $r$ than before.
In particular, in the differences 
between the perturbative predictions
and phenomenological potentials/lattice results,
a linear potential of
order $\Lambda_{\rm QCD}^2 r$ 
at short distances $r \ll \Lambda_{\rm QCD}^{-1}$
was ruled out numerically.

\section{Theoretical Backgrounds}

We would like to understand why
the perturbative predictions
of $E_{\rm tot}(r)$ behave in the way we observed in the
previous sections.
In this section, we review our current theoretical understanding.

\subsection{Perturbative uncertainties: Leading-order renormalon}

The nature of the perturbative series of
$V_{\rm QCD}(r)$ and $E_{\rm tot}(r)$, including their
uncertainties, can be 
understood within the argument based on renormalons.
This argument gives certain estimates of higher-order corrections
in perturbative QCD, and empirically it
gives good estimates even at relatively low orders of perturbative
series.

\begin{figure}[t]\centering
  \psfrag{Etotn}{$E_{\rm tot}^{(n)}(r)$}
  \psfrag{Pole-mass scheme}{Pole-mass scheme}
  \psfrag{MSbar-mass scheme}{$\overline{\rm MS}$-mass scheme}
  \psfrag{NLO}{\raise-3mm\hbox{$\displaystyle \frac{6\pi}{\beta_0\alpha_S}$}}
  \psfrag{LO}{\raise-3mm\hbox{$\displaystyle \frac{2\pi}{\beta_0\alpha_S}$}}
  \psfrag{n}{$n$}
\vspace{7mm}
  \includegraphics[width=8.5cm]{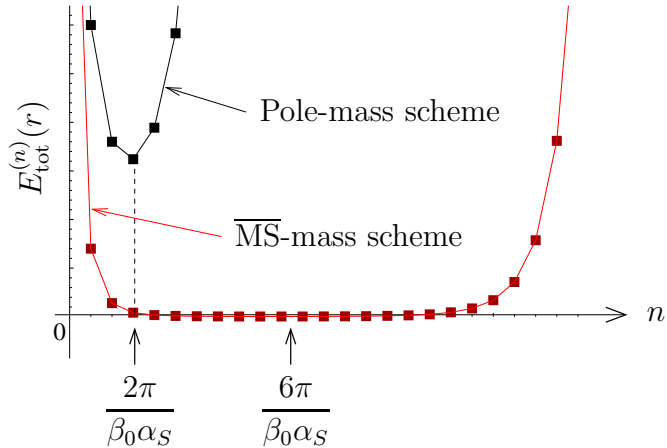}\vspace*{7mm}
\caption{\small
Schematic diagram showing $n$-dependence of 
$|V_{\beta_0}^{(n)}(r)|$ (or
$|E_{\rm tot}^{(n)}(r)|$ in the pole-mass scheme) [black points]
and that of $|E_{\rm tot}^{(n)}(r)|$ in the $\overline{\rm MS}$-mass 
scheme [red points].
\label{Vn}}
\end{figure}

According to the renormalon argument, the perturbative series of
$V_{\rm QCD}(r)$ (or $E_{\rm tot}(r)$ in the pole-mass scheme)
behaves as depicted in Fig.~\ref{Vn} [black points];
see e.g.\ \cite{ren}.
Namely, the series shows apparent convergence up to order
$\alpha_S^{n_0}$ with $n_0 \approx 2\pi/(\beta_0 \alpha_S)$.
Beyond this order, the series diverges.
Due to the divergence
(the series is an asymptotic series), 
there is a limitation to the
achievable accuracy of the perturbative prediction.
It can be estimated by the size of the terms around the
minimum, $O(\alpha_S^{n_0})$;
this gives an uncertainty for
$V_{\rm QCD}(r)$ of order 
$\Lambda_{\rm QCD} \sim 300$~MeV
\cite{al}.

To obtain the above behavior,
we may use, for instance, the ``large-$\beta_0$ approximation'' \cite{bb}
to estimate the higher-order corrections.
The order $\alpha_S^{n+1}$ term 
of $V_{\rm QCD}(r)$ for $n \gg 1$
is estimated as 
\bea
V_{\beta_0}^{(n)}(r)
\sim - \frac{2C_F\alpha_S(\mu)}{\pi} \, \mu \, e^{5/6} \times
\biggl( \frac{\beta_0\alpha_S(\mu)}{2\pi} \biggr)^n \times n! ,
\label{n-th-term}
\eea
where $C_F=4/3$ is a color factor and
$\beta_0 = 11 - 2n_l/3$ is the 1-loop coefficient of the
beta function.
From this expression, it is easy to obtain the behavior of the
series as explained above.

It is important to note that Eq.~(\ref{n-th-term})
is independent of $r$.
Although $V_{\beta_0}^{(n)}(r)$ is dependent on $r$, the leading part
in the large $n$ limit is independent of $r$.
This feature can be verified in Fig.~\ref{figV1L}(a):
\begin{figure}[t]\centering
  \includegraphics[width=16cm]{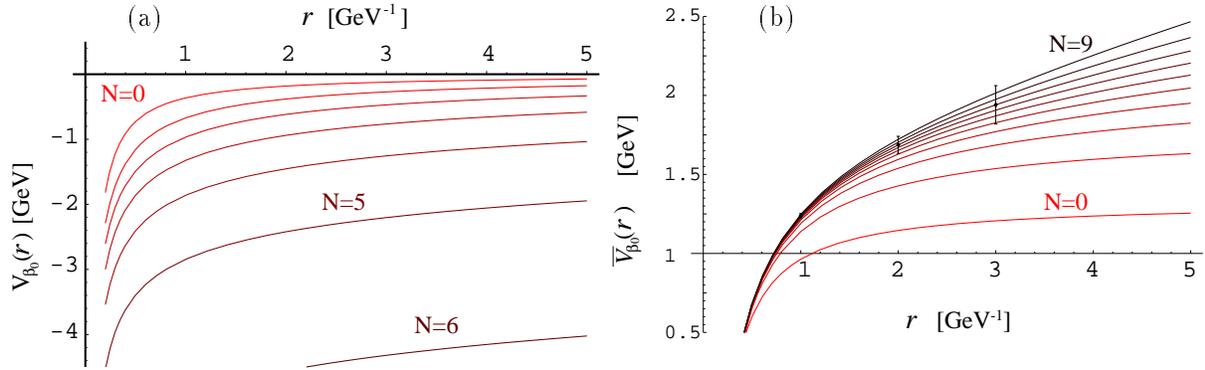}
\caption{\footnotesize
$V_{\beta_0}(r)$ truncated at $O(\alpha_S^{N+1})$ term, 
$\sum_{n=0}^N V_{\beta_0}^{(n)}(r)$.
We set $\mu=2.49$~GeV, $n_l=4$ and $\alpha_S(\mu)=0.273$.
(a) Before cancellation of the leading-order renormalon.
(b) After cancellation of the leading-order renormalon.
Note that the vertical scales are different in two figures.
(The figures are taken from  \cite{Sumino:2001eh}.)
      \label{figV1L}}
\end{figure}
As we include more terms of the perturbative series
of $V_{\rm QCD}(r)$ (in the large-$\beta_0$ approximation),
there are large corrections to the potential, which shift
the potential almost $r$-independently.
Furthermore, one sees no sign of convergence 
of the perturbative series in the
range of $r$ we are interested in.

\subsection{Renormalon cancellation and its implications}

As already stated, 
the leading-order renormalon uncertainty in
the QCD potential gets cancelled in the $Q\bar{Q}$ total energy,
$E_{\rm tot}(r) = 2 m_{\rm pole} + V_{\rm QCD}(r)$,
if we use the $\overline{\rm MS}$ mass.
As seen above, the leading-order renormalon is independent of $r$,
hence, it can be absorbed by a redefinition of the quark mass.
The pole mass is expressed in terms of the  $\overline{\rm MS}$ mass
as a perturbative series, 
$m_{\rm pole} = \overline{m}\, ( 1 + c_1 \alpha_S + c_2 \alpha_S^2 + \cdots )$.
The leading part of this series at large orders takes the same form
as Eq.~(\ref{n-th-term}) but in opposite sign and
with half magnitude \cite{bbb}.
As a result, the leading-order renormalon is cancelled\footnote{
For the complete cancellation of the leading behavior 
Eq.~(\ref{n-th-term}),
one needs to expand $V_{\rm QCD}(r)$ and $m_{\rm pole}$ in the
{\it same} coupling constant $\alpha_S(\mu)$.
This is somewhat involved technically, since usually 
$V_{\rm QCD}(r)$ and $m_{\rm pole}$
are expressed in terms of different coupling constants;
see \cite{Sumino:2001eh,Recksiegel:2001xq,RS-lat}.
}
in $E_{\rm tot}(r)$ 
and its perturbative series behaves as \cite{renormalon}
\bea
E_{\rm tot}^{(n)}(r) \sim {\rm const.}\times r^2 \times
\biggl( \frac{\beta_0\alpha_S(\mu)}{6\pi} \biggr)^n \times n! 
\label{Etotn}
\eea
at large $n$.
As compared to $V_{\beta_0}(r)$,
the series converges faster and up to larger order,
$n \approx 6\pi/(\beta_0\alpha_S)$, 
but beyond this order again the series 
diverges; see Fig.~\ref{Vn}.
An uncertainty of this perturbative series can be estimated from the
size of the terms around the minimum, which gives
$\delta E_{\rm tot}(r) \sim \Lambda_{\rm QCD} \times
(r\Lambda_{\rm QCD})^2$.
Thus, at distances $r \simlt \Lambda_{\rm QCD}^{-1}$,
the uncertainty becomes smaller than the original uncertainty
$\delta V_{\rm QCD}(r) \sim \Lambda_{\rm QCD}$.
In fact, the error bars in Figs.~\ref{comppot}(a,b)
represent $\pm \frac{1}{2}\Lambda^3 r^2$ with $\Lambda=300$~MeV.
The agreement of $E_{\rm tot}(r)$ and phenomenological potentials/lattice
results holds within this uncertainty.

A significant improvement of convergence can be seen by comparing
Figs.~\ref{figV1L}(a) and (b), where higher-order 
corrections are estimated using the large-$\beta_0$ approximation.
Moreover, in Fig.~\ref{figV1L}(b)
we see that the higher-order corrections make the potential 
steeper at large $r$, as compared to the tree-level ($N=0$)
Coulomb potential.
In fact, due to this very feature, agreements between
$E_{\rm tot}(r)$ and phenomenological potentials/lattice results
were observed in Sec.~2.
We may understand the reason as follows.
Let us define a strong coupling constant
$\alpha_F(\mu)$ from the interquark force:
\begin{eqnarray}
F(r) \equiv - \frac{d}{dr} \, V_{\rm QCD}(r) \equiv - C_F \, 
\frac{\alpha_F(1/r)}{r^2}.
\end{eqnarray}
Since the leading-order renormalon in $ V_{\rm QCD}(r)$
is  $r$-independent, it is {\it killed} upon
differentiation by $r$.
$\alpha_F(1/r)$ grows at infrared (IR) due to the running
(the first two coefficients of the $\beta$-function are universal), 
which makes $|F(r)|$
stronger than the Coulomb force at large $r$.
This means that $V_{\rm QCD}(r)$, after cancelling the 
leading-order renormalon,
becomes steeper than the Coulomb potential.
Within perturbative QCD, arguments based on 
$F(r)$ are much more secure than those based on
$V_{\rm QCD}(r)$ in the range of $r$ of our interest.
See \cite{Sumino:2001eh,necco-sommer} for details.

Conventionally, theoretical calculations of the energy
of a heavy $Q\bar{Q}$ boundstate
closely followed that of a QED boundstate such as positronium:
it starts from the natural picture that, when an electron and a 
positron are at rest and 
far apart from each other, they tend to be free particles and the
total energy of the system is given by the sum of the energies of
the two particles (pole masses);
as the electron and positron approach each other, the
energy of the system decreases due to the 
negative potential energy, so that
the total energy of the boundstate is given as 
the sum of the pole masses minus the binding energy.
Applying the same description to the energy of a $Q\bar{Q}$ system,
however, is not natural, because 
when $Q$ and $\bar{Q}$ are far apart from each other,
the free quark picture is not good.
As a result, the perturbative expansion of the boundstate energy
turns out to be poorly convergent, due to the contributions from
IR gluons with wave-lengths of order $\Lambda_{\rm QCD}^{-1}$.
On the other hand, intuitively 
we expect that there should be a way to calculate
the boundstate energy in which the contributions of IR gluons can
be mostly eliminated.
This is because when the boundstate size is sufficiently smaller
than $\Lambda_{\rm QCD}^{-1}$, IR gluons cannot resolve the color
charges of the constituent particles, so that they decouple from
this color-singlet system.
Such a calculation is possible if we use a quark mass
(e.g.\ $\overline{\rm MS}$ mass), 
into which only contributions from short wave-length gluons to the
quark self-energy are absorbed (renormalized).

\begin{figure}[t]
\begin{center}
    \includegraphics[width=14cm]{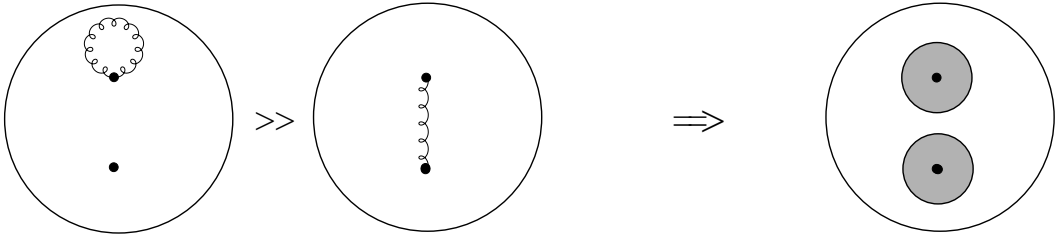}
\end{center}
\caption{\footnotesize
Total energy of a heavy $Q\bar{Q}$ pair is mainly composed of 
(i) the $\overline{\rm MS}$ masses of $Q$ and $\bar{Q}$, and 
(ii) the contributions from gluons with wavelengths
$1/\overline{m} \simlt \lambda_g \simlt r$.
In (ii) the self-energies of $Q$ and $\bar{Q}$
dominate over the potential energy between the two particles.
\label{physpic}}
\end{figure}

According to the renormalon argument 
(in the large-$\beta_0$ approximation),
the QCD potential and the pole mass are roughly given by
\bea
&&
V_{\rm QCD}(r) \approx - \int 
\frac{d^3\vec{q}}{(2\pi)^3} \, e^{i\vec{q}\cdot\vec{r}} \,
C_F \frac{4\pi\alpha_S(q)}{q^2} ,
\label{Vqualit}
\\&&
m_{\rm pole} \approx \overline{m} + \frac{1}{2}
\int\limits_{q\lesssim\overline m}
\frac{d^3\vec{q}}{(2\pi)^3} \, 
C_F \frac{4\pi\alpha_S(q)}{q^2} ,
\eea
where $q=|\vec{q}|$ and
$\alpha_S(q)$ is (essentially) the one-loop running coupling
constant.\footnote{
The large-$\beta_0$ approximation is essentially a resummation
of gluon vacuum polarization, which effectively changes
the coupling constant $\alpha_S(\mu)$ 
to the one-loop running coupling constant,
$\alpha_S(q) = \alpha_S(\mu)/[ 1 - \beta_0\alpha_S(\mu)
\log(\tilde{\mu}/q)/(2\pi) ]$,
where $\tilde{\mu}=e^{5/6}\mu$ in order to account for the non-logarithmic
term of the vacuum polarization.
}
It follows that
qualitatively the $Q\bar{Q}$ total energy can be
expressed as  \cite{bsv1,Sumino:2001eh}\footnote{
IR part of $V_{\rm QCD}(r)$ in Eq.~(\ref{Vqualit}) is cancelled against 
that of $2m_{\rm pole}$, since $e^{i\vec{q}\cdot\vec{r}} \sim 1$
at $q \ll 1/r$.
On the other hand, at $q \gg 1/r$, $e^{i\vec{q}\cdot\vec{r}}$ is 
highly oscillatory
and contributions to $V_{\rm QCD}(r)$ from this region are suppressed.
}
\bea
E_{\rm tot}(r) \simeq 2 \overline{m} + 
\int\limits_{r^{-1}\lesssim q\lesssim\overline m}
\frac{d^3\vec{q}}{(2\pi)^3} \, \,
C_F \frac{4\pi\alpha_S(q)}{q^2} .
\label{app-Etot}
\eea
It shows that the 
energy is mainly composed of
(i) the $\overline{\rm MS}$ masses of $Q$ and $\bar{Q}$, and 
(ii) the self-energies of $Q$ and $\bar{Q}$ originating
from gluons whose wavelengths are shorter than the size of the system
and longer than those absorbed into (i), i.e.\
$1/\overline{m} \simlt \lambda_g \simlt r$.
See Fig.~\ref{physpic}.
Thus, IR gluons ($\lambda_g > r$) 
have decoupled from Eq.~(\ref{app-Etot}).
If $\alpha_S(q)$ were a constant (non-running),
$E_{\rm tot}(r)$ would be a Coulomb potential.
Due to the running of $\alpha_S(q)$, however, there is a
large positive contribution 
in Eq.~(\ref{app-Etot}) as
$r$ increases and gets closer to $\Lambda_{\rm QCD}^{-1}$.
This gives a microscopic description
for how $E_{\rm tot}(r)$ becomes steeper than the Coulomb potential
at large $r$:
It is the rapid growth of the self-energies (ii)
as more IR gluons couple to this system with increasing $r$.
---
Again it stems from the {\it running} of the strong coupling constant.

\subsection{Consistency with operator-product-expansion}

An OPE of the QCD potential
can be performed within the effective field theory 
``potential-NRQCD'' \cite{pinedasoto}.
The dynamical degrees of freedom in this effective theory have
only very soft scale, where typical momenta are $k \ll 1/r$.
Higher momentum scales are integrated out and the effective
interactions are generally non-local in space but local in
time (potentials).
In this framework, IR part of perturbative QCD calculations
is factorized and absorbed into matrix elements of operators.
A factorization scale $\mu_{\rm fac}$ is set to satisfy
$\Lambda_{\rm QCD} \ll \mu_{\rm fac} \ll 1/r$.
Then the total energy of a static $Q\bar{Q}$ pair
is given by\cite{bpsv,Pineda:2002se}
\bea
&&
E_{\rm tot}(r) = 2 m_{\rm pole} + V_{\rm PT}(r;\mu_{\rm fac})
+ \delta E_{\rm US}(r;\mu_{\rm fac}) ,
\label{OPE}
\\ &&
\delta E_{\rm US} = r^2 \times \frac{4\pi\alpha_S}{18}
\int_0^\infty dt \, e^{-it(V_O-V_S)}
\left< \vec{E}^a(t) \phi(t,0)_{ab} \vec{E}^b(0) \right>
(\mu_{\rm fac})
+ \cdots
\eea
The perturbative potential $V_{PT}(r;\mu_{\rm fac})$ is now free from
renormalons.
The renormalons are absorbed into $2 m_{\rm pole}$ and
$\delta E_{\rm US}(r;\mu_{\rm fac})$.
In the leading-order of multipole expansion,
$\delta E_{\rm US}$ is proportional to $r^2$, as shown above.
This is consistent with our observation in Sec.~2.
Namely, the agreement between 
$E_{\rm tot}(r)$ and phenomenological potentials/lattice
results
holds within the ${O}(\Lambda_{\rm QCD}^3r^2)$ renormalon
uncertainty,
while a difference of order $\Lambda_{\rm QCD}^2 r$
between them was ruled out (numerically)
at distances $r \ll \Lambda_{\rm QCD}^{-1}$.
According to OPE, indeed such a (linear)
difference is not allowed theoretically,
since there
exists no operator that can absorb a linear potential.

\section{Pert.\ QCD Potential as a Coulomb+Linear Potential}
\clfn

The observations and theoretical arguments in Secs.~2,3
indicate that the perturbative prediction of
$E_{\rm tot}(r)$ takes a Coulomb+linear form
(up to an $r$-independent constant)
in the range $r \simlt \Lambda_{\rm QCD}^{-1}$, as
represented by a typical phenomenological potential.
We would like to substantiate this claim analytically.
Since the uncertainty of the perturbative prediction is 
of order $r^2$, there is a possibility that $E_{\rm tot}(r)$ 
is predictable up to order $r$ (linear) term 
at $r \simlt \Lambda_{\rm QCD}^{-1}$
within perturbative QCD.

Nevertheless, one immediately comes up with counter arguments:
(1) A finite-order perturbative prediction of
$E_{\rm tot}(r)$ {\it cannot} be expressed in a Coulomb+linear form.
This is because the perturbative QCD potential up to
$O(\alpha_S^N)$ has a form
\bea
V_N(r) = -C_F \frac{\alpha_S(\mu)}{r}
\times \sum_{n=0}^{N-1} \alpha_S(\mu)^n \, P_n(\log \mu r) ,
\label{VN}
\eea
where $P_n(L)$ denotes an $n$-th degree polynomial of $L$.
Therefore, as $r \to \infty$, $V_N(r) \to 0$, i.e.\
the tangent is zero, and there is
no linear potential.
(2) From dimensional analysis, the coefficient of a linear 
potential should be non-analytic in $\alpha_S$, i.e.\
of order $\Lambda_{\rm QCD}^2 \sim \mu^2 \exp[-4\pi/(\beta_0\alpha_S)]$.
Therefore, it should vanish at any order of perturbative expansion.
We will come back to these arguments later.

Since we observed a numerical agreement between 
$E_{\rm tot}(r)$ up to $O(\alpha_S^3)$ and a
Coulomb-plus-linear potential,
it would be interesting to examine how $E_{\rm tot}(r)$
up to $O(\alpha_S^N)$
or $V_N(r)$ will behave at large orders, $N \to \infty$.
For this analysis, we need (a) some estimates of higher-order
terms of $V_N(r)$ and (b) some reasonable scale-fixing prescription.
The latter is needed, since perturbative QCD does not provide any
scale-fixing procedure by itself.
Driven by such interests,
recently we have shown analytically \cite{CplusL} that,
based on the renormalon dominance picture,
$V_N(r)$ quickly ``converges'' to a Coulomb+linear potential
for $N \gg 1$, up to an $O(\Lambda_{\rm QCD}^3 r^2)$ uncertainty.
This was shown (a) by estimating 
the higher-order terms of $V_N(r)$ 
using renormalization group (RG), and 
(b) by adopting a scale-fixing prescription 
based on the renormalon dominance picture.

Let us explain the scale-fixing prescription (b).
According to the renormalon dominance picture,
if we choose a scale $\mu$ such that 
$\alpha_S(\mu) \approx {6\pi}/({\beta_0N})$, around this scale,
$V_N(r)$ (after cancelling the leading-order renomalon) becomes
least $\mu$-dependent and the perturbative series 
becomes most convergent; cf.\ Fig.~\ref{Vn}.
In view of this property, we fix $\mu$ such that\footnote{
Here, we generalize the prescription of \cite{CplusL} by
introducing the parameter $\xi$, cf.\ \cite{BZvAV}.
}
\bea
\alpha_S(\mu) = \frac{6\pi}{\beta_0N} \times \xi
~~~~~~~
(\xi \sim 1).
\label{xi}
\eea
Then we consider $V_N(r)$ for $N \gg 1$ while keeping
$\Lambda_{\overline{\rm MS}}$ finite.

Let us first demonstrate our result in
the simplest case.
We set $\xi = 1$ and
the higher-order terms of
$V_N(r)$ are estimated using the 1-loop running coupling constant
[leading log (LL) approximation]:
\bea
V_{N}(r) = - \int
\frac{d^3\vec{q}}{(2\pi)^3} \, \frac{e^{i \vec{q}\cdot\vec{r}}}
{q^2} \, 4\pi C_F 
\, \left[ \alpha_{1L}(q) \right]_N ,
\eea
where $[\alpha_{1L}(q) ]_N$ denotes the 
perturbative expansion of the 1-loop running coupling constant
truncated at $O(\alpha_S^{N})$:
\bea
\left[ \alpha_{1L}(q)\right]_N 
= \left[ \frac{\alpha_S(\mu)}{1-\frac{\beta_0\alpha_S(\mu)}{2\pi} 
\log\bigl(\frac{\mu}{q}\bigr)} \right]_N
\! \! \! \!
= \alpha_S(\mu)  \, \sum_{n=0}^{N-1}
\Bigl\{ 
{ \frac{\beta_0\alpha_S(\mu)}{2\pi}
\log \bigl( \frac{\mu}{q}\bigr) } 
\Bigr\}^n .
\label{alpha1LN}
\eea
Since the scale $\mu$ and $\alpha_S(\mu)$ are fixed at each $N$
via the relation (\ref{xi}),
$V_N(r)$ is a function of only $r$ and $N$ once
$\Lambda_{\overline{\rm MS}}$ is fixed.

\begin{figure}[t]\centering
  \vspace*{18mm}
  \hspace*{-3cm}
  \includegraphics[width=20.2cm]{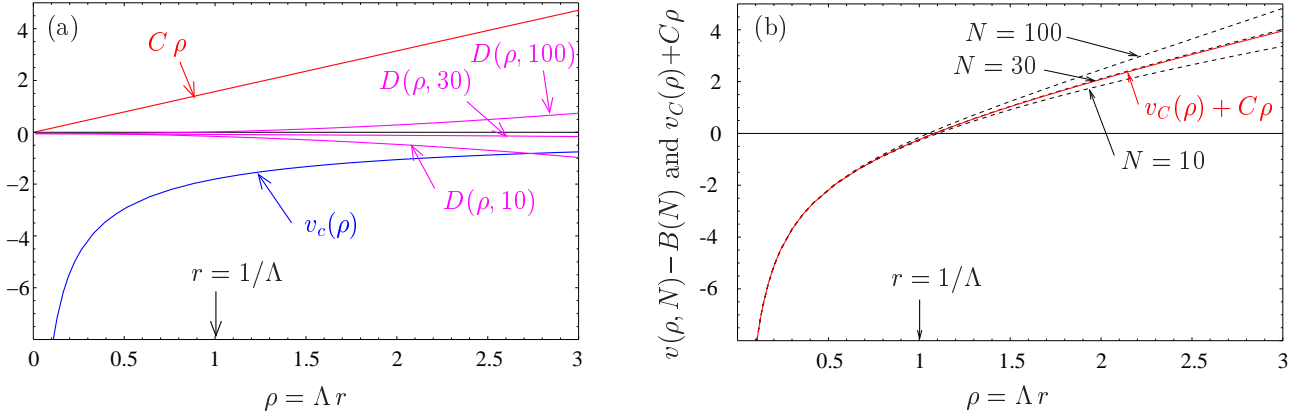}
\caption{\footnotesize
(a) $v_C(\rho)$, $C\rho$ and $D(\rho,N)$ ($N=10,30,100$) vs.\ $\rho$.
(b) Comparison of $v(\rho,N)-B(N)$ ($N=10,30,100$) [dashed black] and the
Coulomb+linear potential $v_C(\rho)+C\rho$ [solid red].
(The latter is hardly distinguishable from the $N=30$ line.)
      \label{decomp}}
\vspace*{5mm}
\end{figure}

In the limit $N \gg 1$, $V_N(r)$ can be
decomposed into four parts corresponding to 
\{$r^{-1}$, $r^0$, $r^1$, $r^2$\} terms
(with log corrections in the $r^{-1}$ and $r^2$ terms):
\bea
&&
V_N(r) 
= \frac{4C_F}{\beta_0} \, \Lambda
\,\, v(\Lambda r,N) ,
\\ &&
v(\rho,N)
=v_C(\rho)+B(N)+C\rho+D(\rho,N)
+(\mbox{terms that vanish as $N\to\infty$}),
~~~
\eea
where we rescaled $r$ and $V_N$ by 
$\Lambda \equiv \mu \, \exp [ - {2\pi}/({\beta_0\alpha_S(\mu)}) ]$.
The ``Coulomb''\footnote{
The ``Coulomb'' part $v_C(\rho)$ contains log
corrections at short-distances and its
short-distance
behavior is consistent with the 1-loop RG equation
for the potential.
}  
[$v_C(\rho)$], linear [$C\rho$] and quadratic [$D(\rho,N)$] parts are
displayed in Fig.~\ref{decomp}(a);
see \cite{CplusL} for the formulas.
The constant [$B(N)$] and quadratic [$D(\rho,N)$] parts are
divergent as $N \to \infty$, while the ``Coulomb''
[$v_C(\rho)$] and
linear [$C\rho$] parts are finite in this limit.
The constant part diverges rapidly, but since it can be 
absorbed by a redefinition of quark mass in the
total energy $E_{\rm tot}(r)$, we will not be concerned with it.
The quadratic part is divergent slowly as
$D \sim (\rho^2 /12)\log N$,
and its size is small for $r \simlt \Lambda^{-1}$ and
$N\simlt 100$ as compared to the Coulomb+linear part;
see Fig.~\ref{decomp}(a).
As a result, when $N$ is increased up to e.g.\ 10--30, 
$v(\rho,N)$ quickly ``converges'' to 
$v_C(\rho)+C\rho$  at $r \simlt \Lambda^{-1}$,
while  it slowly varies as
$(\rho^2 /12)\log N$,
see Fig.~\ref{decomp}(b).
Note that, although $v(\rho,N)$ in this figure have
the form of Eq.~(\ref{VN}),
they approximate well $v_C(\rho)+C\rho$ 
at $r \simlt \Lambda^{-1}$.

If we vary $\xi$ in the scale-fixing prescription Eq.~(\ref{xi}), 
as long as $\xi > 2/3$, we obtain
the same Coulomb+linear potential, $v_C(\rho)+C\rho$,
whereas $B(N)$ and $D(\rho,N)$ change:
for instance,
if $2/3 < \xi < 1$, $D(\rho,N)$ is finite as $N\to\infty$,
$D(\rho) = \rho^P \times (\mbox{log-corr.})$
with $1<P<2$.

We may consider that the constant [$B(N)$] and quadratic
[$D(\rho,N)$] parts represent renormalons in
$V_{\rm QCD}(r)$.
If we perform OPE, these will be (partially) absorbed into
$2m_{\rm pole}$ and $\delta E_{\rm US}$ in Eq.~(\ref{OPE}).
On the other hand, the Coulomb+linear part
[$v_C(\rho)+C\rho$] may be regarded as a genuine 
perturbative part, which cannot be absorbed into 
$2m_{\rm pole}$ and $\delta E_{\rm US}$.

\begin{figure}[t]\centering
  \hspace*{-2.5cm}
  \includegraphics[width=14cm]{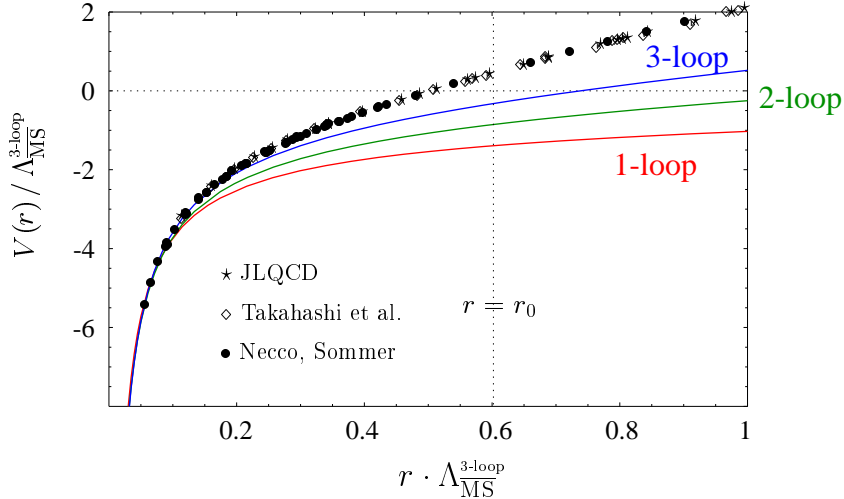}
\caption{\footnotesize
Comparison of the Coulomb+linear potential ($n_l=0$)
with the lattice results in the quenched approximation.
1-, 2- and 3-loop correspond, respectively, to 
LL, NLL and NNLL approximations in the higher-order 
estimates.
      \label{comp-lat}}
\end{figure}

The above result can be systematically improved by incorporating
higher-loop effects [next-to-leading log (NLL), 
next-to-next-to-leading log (NNLL),
$\cdots$]
into the RG estimate of the higher-order terms of
$V_N(r)$.
Fig.~\ref{comp-lat} shows the Coulomb+linear potentials
corresponding to the higher-order estimates by
the 1-loop running (LL), 2-loop running
with 1-loop non-logarithmic term (NLL), and 3-loop running
with 2-loop non-logarithmic term (NNLL).
They are compared with the lattice results.
Since the 3-loop non-logarithmic term is not yet known,
the 3-loop running case represents our current best knowledge.
Up to this order, the Coulomb+linear
potential agrees with the lattice results up to larger $r$
as we increase the order.\footnote{
The NNLL originating from the ultra-soft scale \cite{USLL} hardly
changes the 3-loop running case displayed in Fig.~\ref{comp-lat}
\cite{CplusL}.
}

The coefficient of the
linear potential $\sigma r$ can be expressed analytically.
For the 1- and 2-loop running cases, the expressions read
\bea
&&
\sigma_{\mbox{\scriptsize 1-loop}}= \frac{2\pi C_F}{\beta_0} \,
\Bigl( \Lambda_{\overline{\rm MS}}^{\mbox{\scriptsize 1-loop}} \Bigr)^2 ,
\label{string-tension1}
\\ &&
\sigma_{\mbox{\scriptsize 2-loop}}= \frac{2\pi C_F}{\beta_0} \,
\Bigl( \Lambda_{\overline{\rm MS}}^{\mbox{\scriptsize 2-loop}} \Bigr)^2
\, \frac{e^{-\delta}}{\Gamma (1+\delta)} \,
\biggl[ 1 + \frac{a_1}{\beta_0} \, \delta^{-1-\delta} \, e^\delta \,
\gamma (1+\delta,\delta) \biggr] ,
\label{string-tension2}
\eea
where 
$\delta = \beta_1/\beta_0^2$;
see \cite{CplusL} for details.

Finally, let us comment on the counter arguments given at
the beginning of this section.
(1) $V_N(r)$ ``converges'' to a Coulomb+linear
form for $N \gg 1$.
(See also discussion in \cite{Zakharov}.) 
In fact, already at relatively low orders
and at $r \simlt \Lambda^{-1}$,
$V_N(r)$ approximates the Coulomb+linear potential fairly well.
(2) Consider a perturbative term 
$T=\{ \frac{\beta_0\alpha_S(\mu)}{2\pi} \log (\mu r) \}^N$.
If we substitute the relation (\ref{xi}) and take the limit $N\to \infty$,
it is easy to see that $T \to (\Lambda r)^{3\xi}$.
Thus, if we choose a scale $\mu$  \`a la renormalon dominance picture,
perturbative terms can converge towards $(\Lambda r)^{P}$
with some positive power $P$.
Non-analyticity in $\alpha_S$ enters through the relation (\ref{xi}).

\section{Conclusions}

After cancellation of the leading-order renormalons
by using the $\overline{\rm MS}$ quark mass, perturbative
uncertainty of 
$E_{\rm tot}(r)=2m_{\rm pole} + V_{\rm QCD}(r)$ reduces from
$O(\Lambda_{\rm QCD})$ to $O(\Lambda_{\rm QCD}^3r^2)$
and much more accurate perturbative prediction is obtained
at $r \simlt \Lambda_{\rm QCD}^{-1} \sim 1$~fm.
Consequently we observe the following:
\begin{itemize}
\item
$E_{\rm tot}(r)$ up to $O(\alpha_S^3)$ agrees well with phenomenological
potentials/lattice results within the estimated uncertainty.
\item
New physical picture on the composition of the energy of a
static $Q\bar{Q}$ system or on the interquark force is obtained:
The self-energies of $Q$ and $\bar{Q}$ from gluons with
$\lambda_g < r$ increase rapidly as $r$ increases, which
makes $E_{\rm tot}(r)$ steeper than the Coulomb potential.
\item
Our observations are consistent with OPE within 
potential-NRQCD framework.
\item
Based on the renormalon dominance picture, the perturbative
QCD potential ``converges'' quickly to a Coulomb+linear potential,
which can be computed systematically as we include more terms
in the estimate of higher-order terms via RG.
In particular, the linear potential can be computed analytically
[Eqs.~(\ref{string-tension1},\ref{string-tension2})].
The Coulomb+linear potential agrees with lattice
results up to larger $r$
as we include more terms.
\end{itemize}

The analyses reported in this paper
can be applied to the bottomonium and charmonium
spectroscopy in the frame of potential-NRQCD formalism, in which
the QCD potential is identified with
the leading potential in $1/m$ expansion.
Some applications have already
been done in our recent works \cite{spectroscopy}.
Furthermore, our analyses provide a basis for
(and justification for the error estimates of) the
determination of the bottom and charm quark $\overline{\rm MS}$ masses
from the $\Upsilon(1S)$ and $J/\psi$ energy levels 
computed in perturbative QCD
\cite{bsv1,bsv2}.

\section*{Acknowledgements}
Many of the analyses were done in collaboration with
S.~Recksiegel.
The author is grateful for fruitful discussion.
He also thanks K.~Van Acoleyen and H.~Verschelde
for discussion.

\end{document}